# On-Chip Concentration and Patterning of Biological Cells Using Interplay of Electrical and Thermal Fields

Golak Kunti [a], Tarun Agarwal [b], Anandaroop Bhattacharya [a], Tapas Kumar Maiti [b] and Suman Chakraborty [a]*

[a] Department of Mechanical Engineering, Indian Institute of Technology Kharagpur, Kharagpur, West Bengal - 721302, India

[b] Department of Biotechnology, Indian Institute of Technology Kharagpur, Kharagpur, West Bengal - 721302, India

[*]*E-mail address* of the corresponding author: suman@mech.iitkgp.ernet.in

**Abstract**

We demonstrate a method of concentrating and patterning of biological cells on a chip, exploiting the confluence of electric and thermal fields, without necessitating the use of any external heating or illuminating source.   The technique simply employs two parallel plate electrodes and an insulating layer over the bottom electrode, with a drilled insulating layer for inducing localized variations in the thermal field. A strong induced electric field, in the process, penetrates through the narrow hole and generates highly non-uniform heating, which in turn, results in gradients in electrical properties and induces mobile charges to impose directional fluid flow. The toroidal vortices, induced by secondary electrokinetic forces originating out of temperature-dependent electrical property variations, transport the suspended cells towards a hot-spot site of the chip, for rapid concentrating and patterning into different shaped clusters based on pre-designed conditions, without exceeding safe temperature limits that do not result in damage of thermally labile biological samples. We characterize the efficacy of the cell trapping process for two different biological entities, namely, *Escherichia coli* bacteria and yeast cell. These results may be of profound importance towards developing novel biomedical microdevices for drug discovery, antibiotic resistance assessment and medical diagnostics.

## 1. Introduction

A number of operations, such as trapping, transportation, concentration, re-orientation and sorting of cells has been developed in the clinical and biological fields due to ever-increasing demands for versatile arrangement of living biological entities into desired patterns.[1,2] Patterning of biological objects is the fundamental premise of probing cell-cell interactions, bio-printing, drug development, image-based cell selection etc.[3–9] On the other hand, patterning of cells is widely used to design biosensors for cellular cultivation, and for scaffolds to pattern proteins.[10] Therefore, a technique which can manipulate a large number of cells rapidly and accurately is essential in the area of biomedical engineering. Many techniques have emerged over the years for achieving a desired patterning of cellular entities using various forces, namely mechanical,[11] optical,[12,13] magnetic,[14] electrical [15,16] forces, etc.

Electrically modulated techniques present with many important features of manipulating cells at the micro-scale, since these use the inherent electrical properties of the biological objects without necessitating further labeling. [17,18] Accordingly, they hold promises of patterning cells with high throughput, high speed and high reliability on a microfluidic chip.[18,19] Dielectrophoresis (DEP) has been used widely to manipulate cells in a wide variety of applications showing many advantages, such as induction of low stress on samples, differentiation between particles of different kinds, etc. [20,21] However, DEP uses fixed electrodes which limits its application in many cases.[21] Sometimes, it demands complicated electrodes which need to be fabricated in a small space employing expensive microfabrication methodologies.[22] In addition, the effectiveness of this method is restricted to large particles because dielectrophoresis force is proportional to cubic of particle radius.[23] Moreover, difficulties of adhesion of cells on the surface of the electrodes inhibit continuous operation of cell manipulation for a long time.[24]

In recent years, rapid electrokinetic patterning (REP) has emerged as a noninvasive technique for the manipulation of colloidal particles. [25] This technique uses parallel plate electrodes to generate a uniform alternating electric field. However, REP necessitates high power budget through the deployment of an external highly focused laser source as well as other auxiliary complications, along with a lack of ease.[26] In addition, a large fraction of thermal energy (~85%) gets physically dissipated in the process without any favourable utilization,[27] limiting the deployment of such arrangement in resource-limited settings. Moreover, the system is not self-contained on a chip, since the laser source is used externally.

Here, we introduce a new technique for concentrating and patterning of biological cells on a chip by exploiting the interplay of localized fluid flow, heat transfer and consequent variations in electrical properties. In sharp contrast to other related reported techniques, the present method does not necessitate any external heat source for generating the thermal field. Instead, a simple tweak in the chip design enables the spontaneous inception of a spatially varying temperature distribution in a localized sense. This is essentially achieved by attaching a thin insulating layer with a drilled narrow hole on a bottom electrode on the chip. A non-uniform electric field develops when the electric field penetrates through the hole. As a consequence, electrically induced heating generates a sharp temperature gradient which induces the necessary gradients in electric properties for setting induced charges to fluidic motion.

Here, we explore the above arrangement for the concentration and patterning of biological cells on a microfluidic chip, by deploying the interplay of electrical and thermal fields, albeit, obviating the need of any external heating source. The system is thus self-contained as all the components are embedded on an integrated microfluidic chip. We establish the efficacy of the cell trapping process through demonstration examples on *Escherichia coli* bacteria and yeast cells. The concentration and patterning of cells into different designs establishes its suitability for deployment in biomedical and biochemical assays on a chip.

## 2. Materials and methods

### 2.1. Device fabrication

The device consisted of two parallel plate electrodes, separated by a spacer (Figure 1). Indium-tin-oxide (ITO, thickness 150 nm) coated glass substrates (thickness 1.1 mm) were used as the top and bottom electrodes. The bottom electrode was covered with a layer of cello tape (thickness: $40\,\mu m$) which acted as an insulating layer. Thereafter, a narrow hole (hydraulic diameter $\sim 50\,\mu m$) was drilled at the middle of the insulating layer, taking appropriate precautions to avoid the damage of ITO coating over the glass substrate. Both top and bottom electrodes were then integrated to fabricate the device with double-sided adhesive tape (thickness: $35\,\mu m$) acting as the spacer. The electrodes were connected to an AC function generator (33250A, Agilent) accompanied by a signal voltage amplifier (9200, High Voltage Wide Band Amplifier).

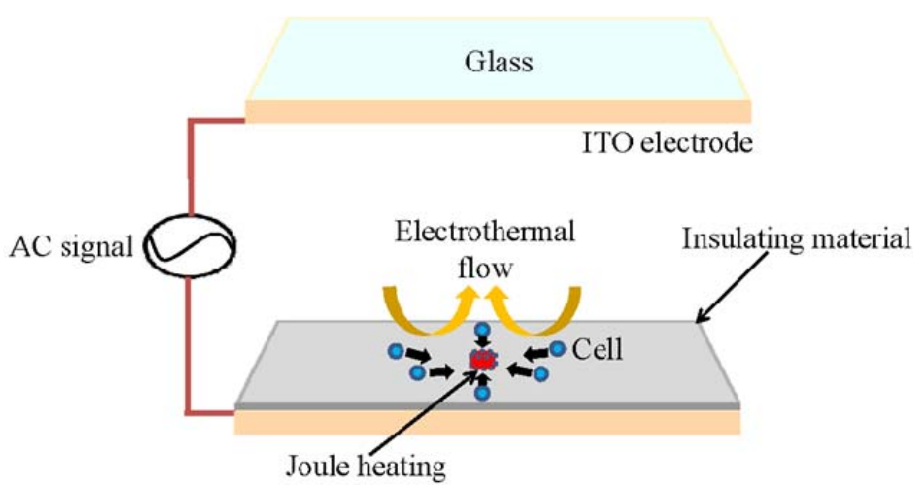

FiG. 1. Schematic illustration of the physical arrangement. AC electric field is applied across the parallel electrodes which results in electrically induced heating at the narrow opening in the bottom electrode (zone of peak temperature). Thermal field and electric field, together, generate toroidal vorties which concentrate the suspended cell at the zone of peak temperature.

## 2.2. Cell culture and preparation

### 2.2.1. Bacterial cells

In the experiments, GFP-expressing bacteria (diameter: 0.8 μm, length: 2-3 μm) were used. For this, the GFP sequence was cloned in pet28a bacterial plasmid vector and transformed in *E. coli*. (BL21 strain) using a standard calcium chloride method. The transformed colonies were grown and selected on Luria agar plates containing kanamycin (400ug/ml). Prior to experimentation, the bacterial cultures were grown in Luria broth (Himedia, India) supplemented with kanamycin (400ug/ml) and IPTG (0.5 mM)) for overnight at 37 °C under mild shaking conditions. Then, the cells were pelleted using centrifugation, washed twice with buffer solution (conductivity, σ ~0.043 S/m, isotonic low conductivity sucrose/dextrose buffer containing 8.3% sucrose and 0.3% dextrose prepared in injection water (SuperAmp®, Otsuka)), and resuspended in the same buffer as per the desired concentration.

### 2.2.2. Yeast cells

To prepare the yeast cells (strain BY4741, diameter: 3-4 μm), the cells were grown in Yeast Extract–Peptone–Dextrose media (Himedia, India) for 24 h at 30 °C under mild shaking conditions. The cells were then harvested, pelleted and washed with phosphate buffered saline (PBS, pH 7.4) and stained with CellTracker™ Red (Life Technologies, prepared in PBS) for 30 min at 30 °C. Thereafter, the cells were washed once with PBS, followed by two washes with buffer (conductivity, σ ~0.043 S/m), and resuspended in the same buffer as per the desired concentration.

## 2.3. Cell patterning methodology

The solution with suspended cells was injected between the electrodes. The electric field was then applied to actuate the cell motion. Induced electrical heating triggered the development of localized temperature gradients, leading to inhomogeneities in the dielectric properties of the buffer solution. These induced electro-thermal forces generated a directional fluid motion, setting the cells into motion in the nearby vicinity. Flow observation was carried out using an inverted microscope (IX71, Olympus) whereas images were captured using a CCD camera (ProgRes MF$^{cool}$) and 10X and 40X objective lens. To characterize the phenomenon of trapping and patterning, micro-particle image velocimetry (μPIV) analysis was carried out. The experiments were performed for a fixed frequency of 500 kHz and a varying voltage of 3.54-9 V (root mean square). A voltage below 3.54 V was insufficient to induce significant cell trapping, while at a voltage above 9 V, electrolytic reactions were observed. Notably, electrolysis reactions were also evident at frequency range below 100kHz.

## 2.4. Theoretical basis

The distribution of electrical potential ($\varphi$) follows: [28,29] $\nabla \cdot (\sigma \nabla \varphi) = 0,$ where $\sigma$ is the electrical conductivity of the medium. The externally applied AC electric field causes heat generation in the bulk fluid domain. This heat source, also known as Joule heat, causes temperature distribution as evident from the energy equation:

$$\rho C_p \frac{\mathrm{D}T}{\mathrm{D}t} = \nabla \cdot \left( k \nabla T \right) + \sigma \left| \mathbf{E} \right|^2, \tag{1}$$

where $C_p$ is the specific heat and $k$ is the thermal conductivity. $\sigma \left| \mathbf{E} \right|^2$ is the Joule heat, $\rho$ is the density of the buffer solution.

The velocity field is characterized by the Navier-Stokes equation augmented by electrical body forces:

$$\rho \frac{\mathrm{D}\mathbf{V}}{\mathrm{D}t} = -\nabla p + \nabla \cdot \left[ \mu \left( \nabla \mathbf{V} + \nabla \mathbf{V}^T \right) \right] + \mathbf{F_E}, \tag{2}$$

Here, $\mathbf{V}$ is the fluid velocity and $\mu$ is the viscosity of the fluid. The time-averaged volumetric force is expressed by [30,31]

$$\mathbf{F_E} = \frac{1}{2} \mathrm{Re} \left[ \frac{\sigma \varepsilon (\alpha - \beta)}{\sigma + i \omega \varepsilon} (\nabla T \cdot \mathbf{E}) \mathbf{E}^* - \frac{1}{2} \varepsilon \alpha \left| \mathbf{E} \right|^2 \nabla T \right], \tag{3}$$

where Re is the real part of []. $\mathbf{E}$ is the electric field and $\mathbf{E}^*$ is the complex conjugate. $\omega$ is the frequency of the AC signal. $\varepsilon$ is the permittivity of the fluid. $\alpha$ and $\beta$ are $(1/\varepsilon)(\partial \varepsilon / \partial T)$ and $(1/\sigma)(\partial \sigma / \partial T)$, respectively. The first term on the right-hand side of Eq. (3) stems from the Coulomb force, whereas, the second term is the dielectric contribution. From the equation, it is also clear that localized temperature gradients cause electrical forcing.

In addition to the electrothermal forces, some other electrical forces, such as dielectrophoretic force, electroosmotic force and electrostatic interactions simultaneously interact, along with van der Waals attraction force near the drilled hole. As a result, a net force prevails in the hotspot area, which effectively acts as a holding force for the cells and causes their aggregation on the bottom electrode. Specifically, DEP which is generated due to the integration between the non-uniform electric field and induced dipole on the polarized cells, dominates over other electrical forces near the hole. At this location, the electric field is highly non-uniform and interaction between the particles and electric field is high.[26] Notably, buoyancy force cannot influence electrothermal velocity for such a configuration where characteristics length scale $l_c \sim 100\,\mu\text{m}$ as they come into play for $l_c > 1000\ \mu\text{m}$.[32]

### 3. Results and discussions

The directional flow from surrounding to the hotspot site, driven by Joule heating-induced (internal heat generation) mechanism, is a multiphysics problem where electric, thermal and flow fields are coupled simultaneously. The buffer solution in the system completes the circuit between the parallel plate electrodes. Upon application of the AC electric field, a non-uniform field gets generated with the maximum field strength at the site of drilled hole (at the bottom electrode), which sharply decreases as the function of distance. Notably, electric field and electrical conductivity together contribute in the Joule heating. However, the conductivity of the domain remains uniform; thus, the thermal field generated resembles the shape of electric field. In the thermal field, a sharp gradient in temperature occurs at the plane just above the insulating layer, with the highest temperature located at the centre of the liquid placed above the hole and the local temperature sharply decreasing with the distance. As a result, temperature-sensitive electrical properties, including conductivity and permittivity, get influenced by the generated temperature gradient, especially in the moieties present around the hotspot. Since the conductivity of the solution maintains the current flow, mobile charges are induced into the system to maintain the current continuity perturbed by the conductivity gradient. On the other hand, mobile charges are also induced by the influence of the gradient of permittivity. Therefore, the gradient of permittivity and conductivity, in tandem, give rise to mobile charges. The flow of the mobile charges, and hence, the flow of solution and the cells, is generated according to variation of temperature. Thus, the cells move from the surrounding to the hotspot site.

One important point to be mentioned here is that the thermal field, which is the source of electrically induced forcing, may affect the cell life and cell activity. It is reported in the literature that the functionalities of the biological samples are spoiled above a temperature rise of 10K.[33] In the present scenario, such limits are never crossed. The bottom line of the biological efficacy of the present approach lies in the fact that it is the temperature gradient that causes the variation in electrical properties to induce mobile charges, and not the absolute temperature by itself. Such a confluence, in principle, has been previously exploited in several other biological applications as well.[34–39]

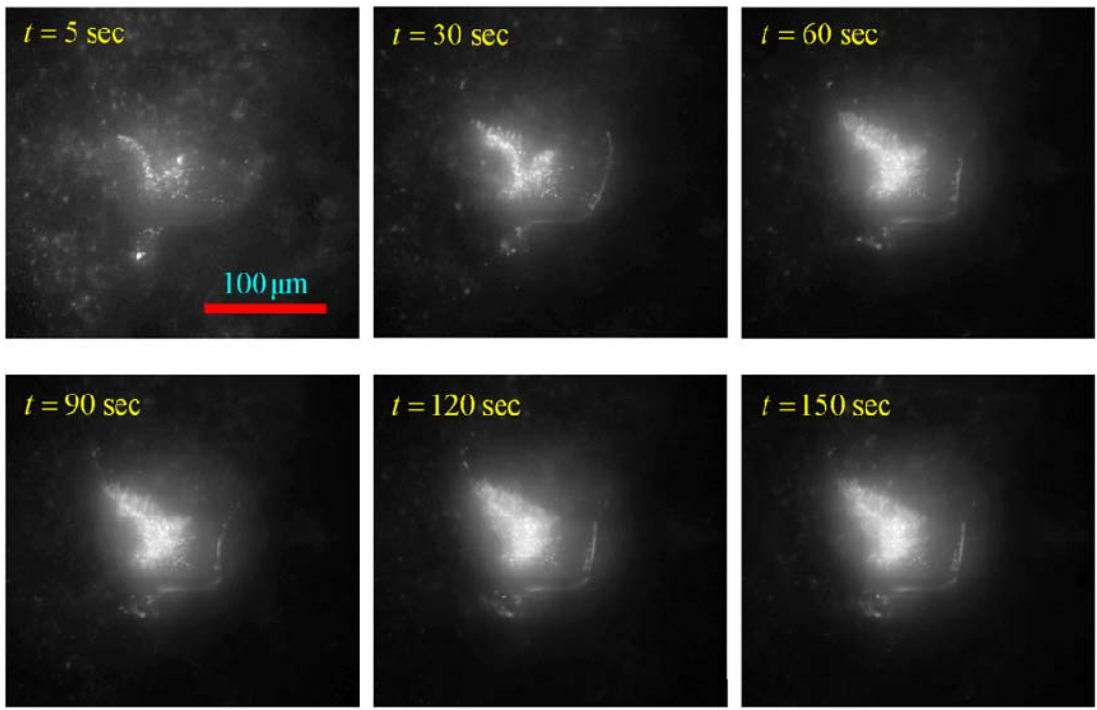

FIG. 2. Time instances ($t = 5\,\text{sec}$, $t = 30\,\text{sec}$, $t = 60\,\text{sec}$, $t = 90\,\text{sec}$, $t = 120\,\text{sec}$ and $t = 150\,\text{sec}$) of images showing *E. coli* concentrating process. After a rapid aggregation in $t = 5-80\,$s, cell aggregation shows gradual accumulation in the time interval of $t = 80-120\,$s, and almost negligible accumulation for the rest of the time. The applied voltage is 7.08 V.

The aggregation of the *E. coli* bacteria is presented in Fig. 2, by highlighting image sequences for an applied voltage of 7.08 V (root mean square). Cell trapping process starts as soon as the electric field is applied. Since the thermal field is produced by exploiting the electric field itself, interplay of electrical and thermal field initiates once the electric field is imposed on the system. This is in stark contrast to laser-induced REP for bioassays that requires electric field and laser beam simultaneously. [18] In the present scenario, electro-thermal flow brings the cells from the surrounding to the hotspot at the plane shown in the figure. Consequently, an outward fluid flow must prevail at the other plane, to maintain overall mass conservation. Despite such continuous flow, cell aggregation and trapping happens by virtue of a net force, primarily dictated by dielectrophoretic effects, which holds the cells accumulated at the trapping spot. The first row of the images shows a rapid change in the number of trapping cells, whereas this change is slow for the second row of images. This is because of a sharp decay of the dielectrophoretic force away from the location of the hole. The region over which the cell accumulation occurs gets rapidly filled by the stack of cluster cells. Eventually, the aggregation rate sharply decreases due to lack of holding force at locations distanced from the drilled hole.

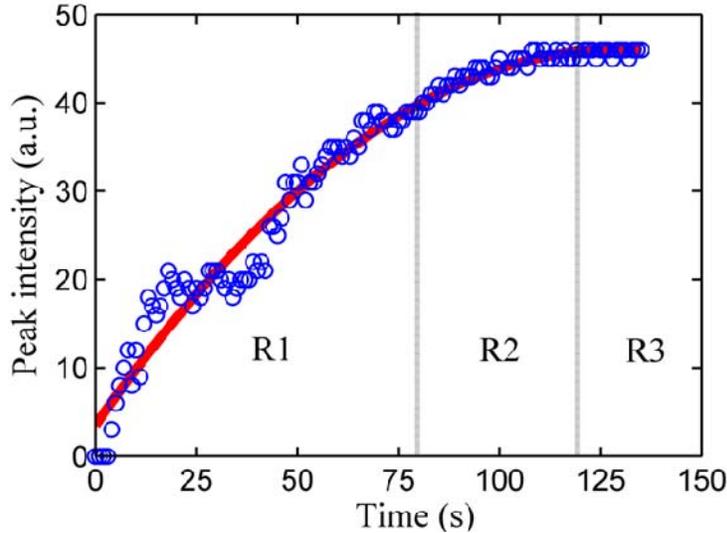

FIG. 3. (a) Increase in maximum fluorescence intensity with time showing rate of aggregation of *E. coli* bacteria. The aggregation process is divided into three regions: R1 (rapid aggregation), R2 (slower aggregation), and R3 (no significant aggregation). The applied voltage is 7.08 V.

We show the time versus fluorescent intensity plot in Fig. 3 for further assessment of the cell concentrating process. The increase in the maximum intensity of the fluorescence with respect to the fluorescent intensity at the initial condition is highlighted in the figure. Maximum fluorescent intensity first increases sharply up to $t \sim 80$ s and then mildly increases up to $t \sim 120$ s. After saturation of the aggregation process, variation of maximum intensity shows a plateau ($t > 120$ s). At the initial stage of the aggregation step, the trapping region is completely blank and all the forces in the system work strongly. However, when the trapping region is filled with cells, the holding force shows the inability to further trap the bioparticles. Further, after significant concentration of the cells, the concentration of the cells in the domain becomes commodious. As a result, the chance of further aggregation becomes less. For clear visualization, the temporal characteristics of intensity are divided into three regions; R1: domain of rapid change in the fluorescent intensity, R2: domain of week change in the fluorescent intensity and R3: domain of negligible change in the fluorescent intensity.

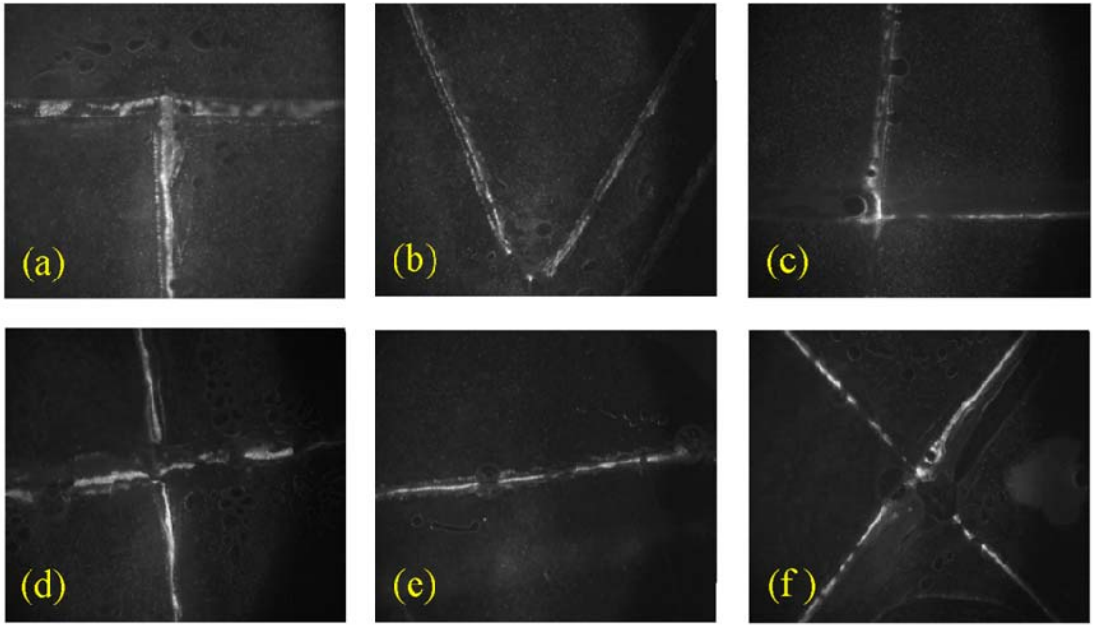

FIG. 4. Different patterns of formation of *E. coli* bacteria: (a) "T", (b) "V", (c) "L", (d) "+", (e) "-", and (f) "×" for $\varphi = 7.08$ V. The widths of the different geometries are in the range of 50-200 μm. The designs of the pattern follow the opened section on the bottom electrode.

Another fascinating aspect of the present technique is the ability to form different cell patterns in a rather elegant manner. *E. coli* drives its body through the rotation of flagella motors attached to the cell membrane. Depending on the direction of rotation, *E. coli* moves along a straight line or rotates at the same place. Alternation of the direction of rotation influences the overall motion. Such motions of *E. coli* were found in our experiments. However, when the electric field is applied, the rotational axis is aligned with the electric field, as a consequence of electro-thermal interactions. Further, different patterns of the *E. coli* are formed by designing different opening structures on the bottom electrode. Previously, it was seen that the cells are concentrated at the location of peak temperature which is observed in the drilled area. Therefore, it is expected that the different structures of the electrically induced heating pattern, induced by tweaking the opening the bottom electrode, can generate different shapes of the clustering cells. Different shapes of the aggregated cells, such as letters "T", "V", "L" and signs "+", "-" "×" could be generated in our experiments (Fig. 4), by resembling the shapes of the pattern formed on the insulating layer. Cell patterning shown here has an important application in tissue engineering for cell culture, specifically during the growth phase of tissue to preserve cell viability.

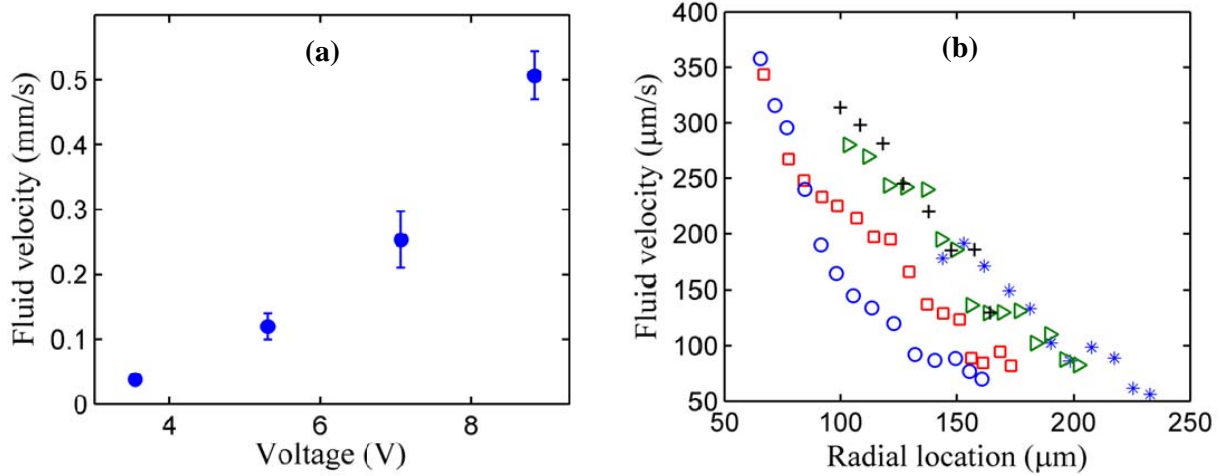

FIG. 5. (a) Plot of voltage versus mean velocity and (b) variation of local velocity with radial distance from the trapping region for $\varphi = 7.08$ V. Mean velocity sharply increases with voltage as electric field strength and Joule heating simultaneously increase with voltage. Due to the rapid drop in the electric field from the trapping region, the local velocity shows a sharp drop near that region.

There are various parameters which directly or indirectly influence the cell concentration and patterning processes. The voltage of the imposed AC signal affects the cell trapping deeply, based on the consideration that the electrically induced heating turns out to be the primary influencing factor in moving the cells towards a target location. In order to reveal the important features of underlying flow physics of electro-thermally modulated cell trapping, the mean velocity data with voltage and spatially varying local velocity are presented in Fig. 5. The mean velocity is evaluated at a distance of 75-100 μm from the trapping region and local velocity is taken at a voltage of 7.08 V. The mean velocity shows a rapid increase in magnitude with increasing voltage. Previously, it was described that the electric field strength and temperature gradients combinedly affect the cell velocity and trend of variations is incremental for both effects. Therefore, it is expected that cell velocity increases rapidly with the applied voltage. To characterize the dependency of the velocity with voltage, it may be noted that ideally, the fluid velocity varies with fourth power of the signal voltage ($V \sim \varphi^4$).[26] This is due to fact that electrothermal force is proportional with square of electric field and varies linearly with temperature gradient, and finally temperature gradient varies with square of the electric field strength ($F_E \sim E^2 \nabla T$, $\nabla T \sim E^2$). However, experimentally obtained data of voltage versus velocity shows an index of 2.93 ($V \sim \varphi^{2.93}$), which is lower than the theoretically predicted index. This deviation from the ideal value may be attributed to the fact that other forces of electrical and non-electrical origin interact with the interplay of electrical and thermal effects near the trapping area. Consequently, the driving electro-thermal force that brings the cells towards the trapping region gets weakened to some extent.

The local velocity first drops sharply near the aggregation spot and then it decays gradually (Fig. 5(b)). As seen before, the fluid velocity strongly depends on the electric field strength. Since the electric field sharply attenuates, the temperature gradient and electro-thermal forces decrease rapidly away from the trapping region. Hence, the local velocity shows a rapid decrease in its magnitude near the trapping region and then decreases slowly.

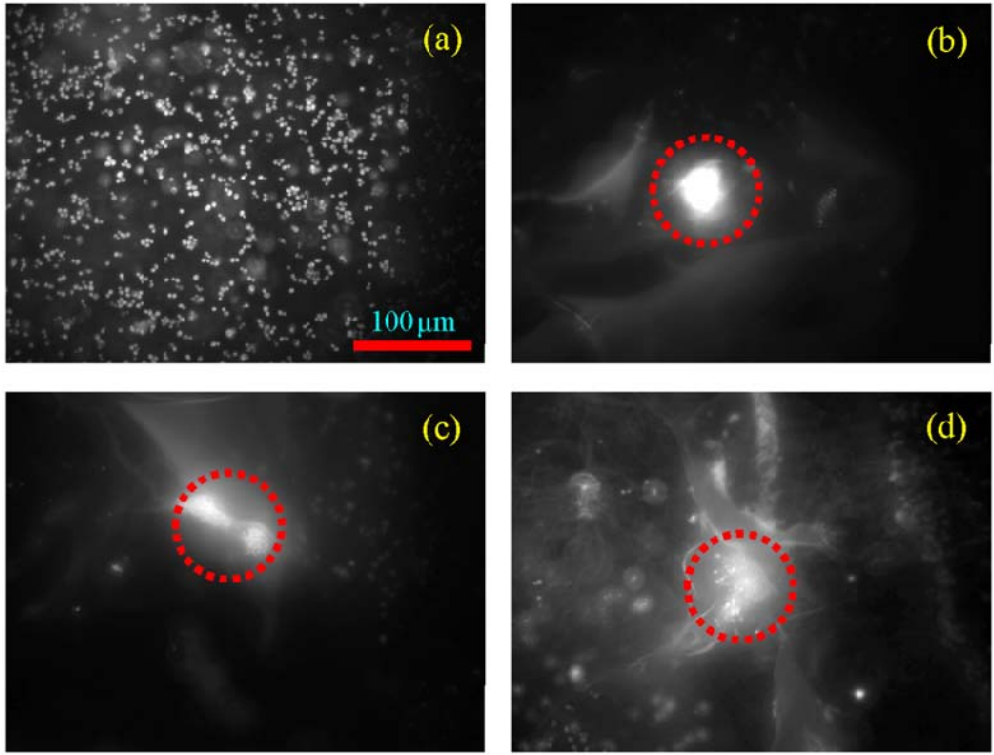

FIG. 6. (a) Distribution of the yeast cells without an electric field. Concentration of the yeast cells for $\varphi = 7.08$ V ((a)-(c)). The hydraulic diameter of the drilled hole is 50-100 $\mu$m . The images are shown at $t = 50$ s from the start of the aggregation process.

To explore the further applicability of our procedure in the biological domain, we have performed experiments on yeast cells as well. We show the aggregation of yeast cells in Fig. 6 for an applied voltage of 7.08 V. Fig. 6(a) depicts the distribution of yeast cells without applying any electric field, whereas Figs.6(a)-(c) highlight the results of cell aggregation in three different sets of experiments. For all cases, we obtain aggregation of yeast. Similarly, the present technique can be applied for wide ranges of bio-particles since the primary influencing factors for such processes turn out to be the forces arising from the interactions between electric and thermal field, and not the interactions between the biological moiety and the electric field.

## 4. Conclusions

We have demonstrated a novel approach of continuous cell concentration and patterning that exploits the interplay between the non-uniform electric field and non-uniform thermal field, without necessitating the deployment of an external heating source. Using an alternating current electric field along with a simple design of the microfluidic chip, the technique enables trapping of different biological entities, such as *Escherichia coli* bacterial cell and yeast cell with high throughput, in an efficient way. Further, the method does not necessitate the pre-treatment of the biosamples. The electrokinetic processor developed in the present study further holds the ability to enhance reproducibility and response of biosensors by allowing simple and fast cell analysis. With several important advantages, such as operation at low power budget, high speed, high-accuracy and yet simplicity in design, the present technique can be applied in different biological contexts ranging from tissue engineering, drug screening, microarray to cell studies. This, in conjunction with other explored avenues such as fluid pumping, [40,41] mixing of analytes, [42,43] controlling two-phase flow[44,45], etc., open of new vistas in biomedical and microfluidic technology.

**Acknowledgement**

This research has been supported by Indian Institute of Technology Kharagpur, India [Sanction Letter no.: IIT/SRIC/ATDC/CEM/2013-14/118, dated 19.12.2013]. SC gratefully acknowledges Department of Science and Technology, Government of India, for Sir J. C. Bose National Fellowship.

**References**

[1] R. Edmondson, J.J. Broglie, A.F. Adcock, and L. Yang, Assay and Drug Development Technologies **12**, 207 (2014).

[2] N. Kramer, A. Walzl, C. Unger, M. Rosner, G. Krupitza, M. Hengstschläger, and H. Dolznig, Mutation Research/Reviews in Mutation Research **752**, 10 (2013).

[3] J. Nilsson, M. Evander, B. Hammarström, and T. Laurell, Analytica Chimica Acta **649**, 141 (2009).

[4] B. Guillotin and F. Guillemot, Trends in Biotechnology **29**, 183 (2011).

[5] L. Kang, B.G. Chung, R. Langer, and A. Khademhosseini, Drug Discovery Today **13**, 1 (2008).

[6] B.M. Taff and J. Voldman, Analytical Chemistry **77**, 7976 (2005).

[7] S. Kar, T.K. Maiti, and S. Chakraborty, Analyst **140**, 6473 (2015).

[8] K. Chaudhury, U. Ghosh, and S. Chakraborty, International Journal of Heat and Mass Transfer **67**, 1083 (2013).

[9] S. Chakraborty, Lab on a Chip **5**, 421 (2005).

[10] K. Ino, H. Shiku, F. Ozawa, T. Yasukawa, and T. Matsue, Biotechnology and Bioengineering **104**, 709 (2009).

[11] J.R. Rettig and A. Folch, Analytical Chemistry **77**, 5628 (2005).

[12] F. Arai, C. Ng, H. Maruyama, A. Ichikawa, H. El-Shimy, and T. Fukuda, Lab on a Chip **5**, 1399 (2005).

[13] X. Wang, S. Chen, M. Kong, Z. Wang, K.D. Costa, R.A. Li, and D. Sun, Lab on a Chip **11**, 3656 (2011).

[14] N. Pamme and A. Manz, Analytical Chemistry **76**, 7250 (2004).

[15] X. Hu, P.H. Bessette, J. Qian, C.D. Meinhart, P.S. Daugherty, and H.T. Soh, Proceedings of the National Academy of Sciences **102**, 15757 (2005).

[16] I. Cheng, V.E. Froude, Y. Zhu, H. Chang, and H. Chang, Lab on a Chip **9**, 3193 (2009).

[17] B.A. Nestor, E. Samiei, R. Samanipour, A. Gupta, A. den Berg, M.D. de Leon Derby, Z. Wang, H.R. Nejad, K. Kim, and M. Hoorfar, RSC Advances **6**, 57409 (2016).

[18] J.-S. Kwon, S.P. Ravindranath, A. Kumar, J. Irudayaraj, and S.T. Wereley, Lab on a Chip **12**, 4955 (2012).

[19] L.-C. Hsiung, C.-H. Yang, C.-L. Chiu, C.-L. Chen, Y. Wang, H. Lee, J.-Y. Cheng, M.-C. Ho, and A.M. Wo, Biosensors and Bioelectronics **24**, 869 (2008).

[20] C. Iliescu, G. Xu, W.H. Tong, F. Yu, C.M. B\ualan, G. Tresset, and H. Yu, Microfluidics and Nanofluidics **19**, 363 (2015).

[21] K.-C. Wang, A. Kumar, S.J. Williams, N.G. Green, K.C. Kim, and H.-S. Chuang, Lab Chip **14**, 3958 (2014).

[22] H. Morgan, M.P. Hughes, and N.G. Green, Biophysical Journal **77**, 516 (1999).

[23] V. Velasco and S.J. Williams, Journal of Colloid and Interface Science **394**, 598 (2013).

[24] I.R. Perch-Nielsen, D.D. Bang, C.R. Poulsen, J. El-Ali, and A. Wolff, Lab on a Chip **3**, 212 (2003).

[25] S.J. Williams, A. Kumar, and S.T. Wereley, Lab on a Chip **8**, 1879 (2008).

[26] G. Kunti, J. Dhar, A. Bhattacharya, and S. Chakraborty, Biomicrofluidics **13**, 014113 (2019).

[27] S.J. Williams, A. Kumar, G. Green, and S.T. Wereley, Nanoscale **1**, 133 (2009).

[28] Castellanos A, *Electrohydrodynamics* (Springer, New York, 1998).

[29] F.J. Hong, F. Bai, and P. Cheng, Microfluidics and Nanofluidics **13**, 411 (2012).

[30] N.G. Green, A. Ramos, A. Gonzalez, A. Castellanos, and H. Morgan, Journal of Electrostatics **53**, 71 (2001).

[31] A. Ramos, H. Morgan, N.G. Green, and A. Castellanos, Journal of Physics D: Applied Physics **31**, 2338 (1998).

[32] A. González, A. Ramos, H. Morgan, N.G. Green, and A. Castellanos, J Fluid Mech **564**, 415 (2006).

[33] S. Lim, H. Park, E. Choi, and J. Kim, Key Engineering Materials **343**, 537 (2007).

[34] J. Wu, M. Lian, and K. Yang, Applied Physics Letters **90**, 234103 (2007).

[35] Q. Yuan, K. Yang, and J. Wu, Microfluidics and Nanofluidics **16**, 167 (2014).

[36] Q. Yuan and J. Wu, Biomedical Microdevices **15**, 125 (2013).

[37] R.H. Vafaie, H.B. Ghavifekr, H. Van Lintel, J. Brugger, and P. Renaud, Electrophoresis **37**, 719 (2016).

[38] Q. Lang, Y. Ren, D. Hobson, Y. Tao, L. Hou, Y. Jia, Q. Hu, J. Liu, X. Zhao, and H. Jiang, Biomicrofluidics **10**, 064102 (2016).

[39] J. Gao, M.L.Y. Sin, T. Liu, V. Gau, J.C. Liao, and P.K. Wong, Lab on a Chip **11**, 1770 (2011).

[40] A. Salari, M. Navi, and C. Dalton, Biomicrofluidics **9**, 014113 (2015).

[41] G. Kunti, J. Dhar, A. Bhattacharya, and S. Chakraborty, Journal of Applied Physics **123**, 244901 (2018).

[42] J.J. Feng, S. Krishnamoorthy, and S. Sundaram, Biomicrofluidics **1**, 024102 (2007).

[43] G. Kunti, A. Bhattacharya, and S. Chakraborty, Electrophoresis **38**, 1310 (2017).

[44] G. Kunti, A. Bhattacharya, and S. Chakraborty, Physics of Fluids **31**, 32002 (2019).

[45] G. Kunti, A. Bhattacharya, and S. Chakraborty, Soft Matter **13**, 6377 (2017).